\documentclass{iopjournal}

\usepackage{physics}
\usepackage{float}
\usepackage{xcolor}
\usepackage{soul}
\usepackage[normalem]{ulem}
\usepackage[mathscr]{euscript}
\usepackage{bm}
\newcommand{\bkfa}{Ba$_{1-x}$K$_x$Fe$_2$As$_2$ }
\usepackage{dsfont}
\usepackage{float}

\newcommand{\ii}{\mathrm{i}}
\newcommand{\ee}{\mathrm{e}}
\renewcommand{\Vec}[1]{\bm{#1}}
\newcommand{\hc}{{\rm{H.c.}}}

\newcommand{\B}{{\rm{B}}}

\newcommand{\figref}{FIG.~\ref}

\renewcommand{\eqref}[1]{{Eq.~(\ref{#1})}}

\begin{document}

\articletype{Paper} 

\title{Microscopic properties of  fractional vortices and domain walls in three-band $s+is$ superconductors}

\author{Igor Timoshuk$^*$\orcid{0000-0002-9317-5154} and Egor Babaev\orcid{0000-0001-7593-4543}}

\affil{Department of Physics, The Royal Institute of Technology, Stockholm SE-10691, Sweden}

\affil{Wallenberg Initiative Materials Science for Sustainability, Department of Physics, The Royal Institute of Technology, Stockholm SE-10691, Sweden}

\affil{$^*$Author to whom any correspondence should be addressed.}

\email{timoshuk@kth.se}


\begin{abstract}
Several experimental observations of objects carrying fractional flux quanta in superconductors were recently reported. Here, we provide microscopic solutions for vortices carrying a variable fraction of magnetic flux quantum and domain walls in a three-band $s+is$ superconductor and investigate their properties. We obtain solutions  in a fully self-consistent treatment of a microscopic three-band Bogoliubov-de-Gennes model. This demonstrates the 
characteristic patterns for the magnetic field distribution. 
The microscopic formalism allows for calculating  tunneling conductance that may be used to distinguish fractional vortices from conventional single flux quanta vortices in Scanning Tunneling Microscopy.   
\end{abstract}

\section{Introduction}

The flux quantization is a fundamental property of conventional superconductors \cite{london1948problem,london1950superfluids}.
In London's argument,  the flux quantization is a consequence of the quantization of the circulation of the $2 \pi$-periodic phase variable, called the phase winding $\oint \nabla \theta = 2\pi N$ where $N$ is an integer. At a deeper level, the existence of the phase comes as a consequence of the concept that a superconductor is a state of matter that spontaneously breaks $U(1)$ gauge symmetry and, hence, is
described by a complex order parameter field \cite{Ginzburg1950,Gorkov1959}. These circumstances ensure that in an infinite sample, there are no stable solutions violating the quantization \footnote{Note that no such penalty exists in mesoscopic samples. Also, layered systems and systems with phase jumps at grain boundaries and weak links have geometric effects that can affect enclosed flux; here, we do not discuss such geometric effects}.
Similar symmetry-based arguments were used to predict that chiral $p$-wave superconductors allow half-quantum vortices \cite{sigrist1991phenomenological,volovik2000monopoles}. 
A different proposal was made in \cite{Babaev2002vortices}, where it was suggested that, in contrast to symmetry-based arguments, vorticity could exist in a gap field of an individual band of a multiband superconductor. In that case, a vortex can carry an arbitrary fraction of a magnetic flux quantum, coined fractional vortices (recently, such vortices were also termed unquantized vortices).
A gap field in an individual band is not an order parameter in the sense of Landau's theory.
Namely, adhering to symmetry-based Landau theory would imply only one order parameter since only one symmetry is broken. Therefore, it would imply the presence of only one phase field, precluding the existence of multiple types of vortices including ones that carry an arbitrary fraction of the magnetic flux quantum.
Notably, in the conventional argument,  a classical field --- the order parameter --should describe all macroscopic aspects of motion regardless of microscopic detail; the number of bands is one of such ostensibly inconsequential microscopic details. A three-band superconductor can break more than $U(1)$ symmetry: when there is a two-fold degenerate phase-difference locking, such a system breaks $U(1)\times Z_2$. In that case the problem is similar: the symmetry only guarantees two types of topological defects, vortices and domain walls, while the possibility of having phase windings in three different bands, exceeds the number of broken symmetries and would not be consistent with the standard symmetry-based arguments.

On the other hand, a classical field theory concept is more general than an order parameter concept and does not necessarily require a set of broken symmetries.
The initial constructions of ``fractional" vortex in \cite{Babaev2002vortices} in a two-band model with single broken symmetry relied on several assumptions. Firstly, 
in the presence of the Josephson coupling between components of a two-component model, there are no well-defined multiple ``Mexican hat" effective potentials with multiple degenerate valleys. Hence, one cannot define two phase variables using the usual symmetry principles. Nonetheless, it was argued, at the level of phenomenological models in \cite{Babaev2002vortices}, that there can be a ``deformed" and ``non-degenerate" valley in the landscape of a free energy functional that still allows the introduction of two classical phase-like variables and sustains energetically stable windings in the phase variables. 
The recent experiment  \cite{iguchi2023superconducting} reported the observation of vortices carrying a temperature-dependent fraction of the flux quantum in \bkfa.
Notably, the experiments demonstrated that the objects with fractional flux (i) could be created in various locations of the samples, giving consistent flux fractions irrespective of their position, (ii) vortices are mobile, and their position can be manipulated, (iii) it was possible to create fractional and integer vortices at the same position, which proves that the observations are not artifacts of some crystal defects. Recently, vortex core fractionalization was observed in a related material, KFe$_2$As$_2$ \cite{newfract}. 

Besides the phenomenon being interesting on its own, the fractional vortices should exhibit anyonic statistics. According to the earlier work \cite{wilczek1982quantum}, a composite object of a non-integer flux tube and an electron is an anyon \cite{wilczek1982quantum,nayak2008non,shapere1989geometric,leinaas1977theory}. Hence, an electron occupying the core state of the fractional vortex physically realizes this object in a superconductor. Consequently, electrons in fractional vortex cores should obey anyon statistics. 

The material \bkfa where fractional vortices were observed is a multiband, spin-singlet superconductor.
At the doping value $x\approx 0.77$, it breaks the time-reversal symmetry \cite{Grinenko2017superconductivity,Grinenko2018superconductivity}.
The time-reversal symmetry is broken above the superconducting phase transition \cite{grinenkonew,shipulin2022calorimetric}, due to fluctuation effects \cite{Bojesen2013time,Bojesen2014phase,maccari2022effects}.
This guarantees that the order parameter has at least two nonzero components when the system transitions to the diamagnetic state. However, the situation is more complex, even with multiple broken symmetries in multiband superconductors.
By the same argument as discussed above,
more than two phase-like variables are possible in \bkfa, despite only two broken symmetries, which implies possibly more than two kinds of fractional vortices \cite{garaud2014domain,Babaev2002vortices,benfenati2023demoisntration}. The presence of nontrivial phase differences between these components makes complex conjugation not equivalent to trivial phase rotation, introducing additional $Z_2$ symmetry.
It is still a subject of current experimental research how many types of fractional vortices can exist in \bkfa. 

Since the existence of ``fractional vortices" cannot be guaranteed by symmetry, and instead, one relies on interaction to produce a particular energy landscape, a detailed microscopic investigation of an isolated vortex solution is warranted. 
This paper presents such solutions in a fully self-consistent microscopic model, including intercomponent gauge-field coupling. A similar solution was obtained in the Bogoliubov - de Gennes (BdG) model in Ref.\cite{iguchi2023superconducting}, but only the magnetic field distribution patterns were analyzed.

The rest of the paper is organized as follows: in Sec. \ref{sec:model} we formulate the model, introducing the studied mean-field Hamiltonian. Sec. \ref{sec:results} starts with comparing the fractional vortices in the systems with and without coupling between different superconducting bands. Sec. \ref{subsec:dw} describes the properties of the domain walls, accompanying fractional vortices in the systems with nonzero interband coupling. These domain walls, in principle, may be used to move the fractional vortices. Sections \ref{subsec:tunn_cond} and \ref{subsec:B_field} describe the observable properties, tunneling conductance, and magnetic field distribution, respectively, characterizing fractional vortices. In Sec. \ref{sec:core_states}, we show that a combination of a fractional vortex and an electron localized in its core obeys anyon statistics. Sec. \ref{sec:concl} contains our conclusions. 

\section{Model}\label{sec:model}
We focus on a three-band model where the interband Josephson coupling ensures that the model has only a single local $U(1)$ symmetry spontaneously broken.
We define the model on a two-dimensional square lattice, described by the microscopic Hamiltonian %
\begin{equation} \label{eq: full Hubbard hamiltonian}
\begin{aligned}
        H = - \sum_{\alpha\sigma}\sum_{<ij>}\exp{\ii q A_{ij} }c^\dagger_{i\sigma\alpha}c_{j\sigma\alpha}  -\sum_{i\alpha\beta}V_{\alpha\beta} c_{i\uparrow\alpha}^\dagger c_{i\downarrow\alpha}^\dagger c_{i\downarrow\beta} c_{i\uparrow\beta}\,.
\end{aligned}
\end{equation}

Here $<ij>$ denotes nearest neighbor pairs, $c_{i \sigma \alpha}$ is the fermionic annihilation operator at position $i$, with spin $\sigma$ ($\sigma \in \lbrace \uparrow, \downarrow \rbrace$) and in band $\alpha$ ($\alpha \in \lbrace 1,2,3 \rbrace$), and phase factor $\exp{\ii q A_{ij}}$ accounts for interaction with the magnetic vector potential $ A_{ij} = \int_j^i \Vec{A} \cdot \dd{\Vec{\ell}}$ through Peierls substitution \cite{Peierl,feynman2011mainly}. The dimensional constants are absorbed into the dimensionless charge $q=\frac{e}{\hbar}a\sqrt{t\mu_0/L_Z}$, where $a$ is the lattice spacing, $t$ is the nearest-neighbor hopping parameter, $e$ is the electronic charge, $\mu_0$ is vacuum permeability, and $L_z$ is the perpendicular length scale associated with the magnetic field.
The quartic interaction term, defined by $V_{\alpha \beta} = V_{\beta \alpha}^*$, allows Cooper pairs to form and tunnel between bands. 
Inter-band coupling dictates no extra $U(1)$ degeneracies. We consider repulsive interband Josephson coupling so that the model also breaks time-reversal symmetry \cite{Stanev2010,Boeker2017} to make the connection with \bkfa \cite{Grinenko2017superconductivity,Grinenko2018superconductivity}. The model has two broken symmetries but three components. Hence, three types of fractional vortices in this model cannot be justified by the standard ``ground state manifold"-based topological classification of defects.

By performing the mean field approximation in the Cooper channel, we obtain the mean-field Hamiltonian
\begin{gather}
\begin{gathered}\label{eq: mfHamiltonian}
    \mathcal{H}=  -  \sum_{\sigma\alpha}\sum_{<ij>} \exp{\ii q A_{ij} } c^\dagger_{i \sigma \alpha } c_{j\sigma \alpha} 
     + \sum_{i \alpha} \left( \Delta_{i \alpha} c^\dagger_{\uparrow i \alpha} c^\dagger_{\downarrow i \alpha} +  \hc \right) + \frac{1}{2} \sum_{\rm{plaquettes}}B_z^2 \,,
\end{gathered} \\
    \Delta_{i \alpha} = \sum_\beta V_{\alpha \beta} \expval{c_{\uparrow i \beta} c_{\downarrow i \beta}} \,,\label{eq: selfConsistency}
\end{gather}
where $\hc$ denotes Hermitian conjugation. Discrete version of Maxwell's equation $\nabla \times \nabla \times \Vec{A} = \Vec{J}$   determines the connection between $A_{ij}$ and charge current $ J_{ij}$ 
\begin{equation}\label{eq:curr}
    J_{ij} = -2q \sum_{\alpha\sigma} \Im{ \expval{c_{i\sigma\alpha}^\dagger c_{j \sigma \alpha}} \exp{\ii q A_{ij} }}
\end{equation}
in the form
\begin{equation}\label{eq:vp_update}
    \expval{\frac{\partial\mathcal{H}}{\partial A_{ij}}}=\frac{1}{2}\frac{\partial}{\partial A_{ij}}\sum_{\mathrm{plaquettes}}B^2-J_{ij}=0,
\end{equation}
where the magnetic field $\Vec{B}=\nabla \times \Vec{A}$ is defined on plaquettes. Here, we always assume that the magnetic field outside the sample is zero.

The free energy for the system \eqref{eq: mfHamiltonian} can be calculated as 

\begin{equation}\label{eq:Fsum}
\begin{gathered}
    F_H = \sum_{i}\Vec{\Delta}_i^\dagger V^{-1}\Vec{\Delta}_i -k_\B T\Tr\ln\qty(\ee^{-\beta \mathcal{H}}+1) 
    + \frac{1}{2}\sum_{\mathrm{plaquettes}} B^2.
\end{gathered}
\end{equation} 

Using the iteration scheme, described in \cite{benfenati2023demoisntration}, the solution for equations \eqref{eq: selfConsistency}, \eqref{eq:curr}, along with the Maxwell equation, is obtained. Each iteration is performed in two steps. First, we calculate the thermal averages in \eqref{eq: selfConsistency}, \eqref{eq:curr} using the approximate Chebyshev spectral expansion method. Secondly, updated $\Delta_{i\alpha}$ and $J_{ij}$ are used to perform gradient decent iteration for vector potential $A_{ij}$ to minimize $\langle\mathcal{H}\rangle$ using  (\ref{eq:vp_update}). The self-consistent iteration procedure stops when the convergence criteria $\abs{\delta p/\left(p +\epsilon\right)} < \varepsilon$ is achieved for each of the parameters $\Delta_1$, $\Delta_2$, $\Delta_3$ and $A$ simultaneously.

The tunneling conductance in the system with externally applied voltage $U$ probes the density of states at the energy $E=E_F-qU$. Thus, after obtaining a self-consistent solution for $\Delta_\alpha$ and $A$, the tunneling conductance is expressed as
\begin{equation}\label{eq:tunn_cond}
\begin{gathered}
        \frac{\partial I_{i}\left(U\right)}{\partial U }\propto \sum_{\alpha n}\left[\left|u_{i \alpha n}\right|^2 f'\left(E_{\alpha n} - q U\right) + \left|v_{i \alpha n}\right|^2f'\left(E_{\alpha n} + q U\right)\right]\,, 
\end{gathered}
\end{equation}
where $f'$ is a derivative of the Fermi-function, $i$ indicates the lattice point, $\alpha$ and $n$ denote eigenstate of the system in the band $\alpha$, with energy $E_{\alpha n}$, electronic and hole distributions $u_{i \alpha n}$ and $v_{i \alpha n}$ respectively, and $U
$ is applied voltage \cite{gygi1990tun, gygi1990ang, gygi1991elstr}.

\section{Results}\label{sec:results}

This paper analyzes the solutions for square systems with linear sizes up to $L=64$ nodes. We consider system with symmetric intraband interaction $V_{11}=V_{22}=V_{33} = 2.4$ and negative interband coupling $V_{12}=V_{13}=V_{23}=-0.6$ with fixed dimensionless charge $q=0.25$ and various temperatures. We also investigate case of slightly non-symmetric intraband interaction $V_{11}=2.5$, $V_{22}=V_{33} = 2.4$, $V_{12}=V_{13}=V_{23}=-u$ for $u=0.6,\, 0.3,\, 0.1$. $\epsilon=10^{-8}$, $\varepsilon=10^{-6}$ are selected as convergence parameters for all simulations.

We start the iteration procedure with an initial guess for the superconducting gaps \eqref{eq: selfConsistency} and vector potential $A$  and converge to a self-consistent solution. Here, we consider four types of initial conditions, yielding different objects. To obtain a uniform solution without any vortex or domain wall, we set $\Delta_{i1}=0.1$, $\Delta_{i2}=0.1\cdot e^{2\pi i/3}$, and $\Delta_{i3}=0.1\cdot e^{-2\pi i/3}$. A domain wall is generated by the same initial guess in the left half of the sample, and with $\Delta_{i2}=0.1\cdot e^{-2\pi i/3}$, $\Delta_{i3}=0.1\cdot e^{2\pi i/3}$ in the right one. Both configurations have optimal phase locking. However, these solutions have different phase ordering: $\arg \Delta_{i1}<\arg \Delta_{i2}<\arg \Delta_{i3}$ on the left and $\arg \Delta_{i1}<\arg \Delta_{i3}<\arg \Delta_{i2}$ on the right. So, there are two different domains in this system.  A fractional vortex is generated by setting $\Delta_{i1}=0.1\cdot\tanh\left(20r_i/L\right)\cdot e^{i\varphi_i}$, $\Delta_{i2}=0.1$, and $\Delta_{i3}=-0.1$, where $r_i$ and $\varphi_i$ are polar coordinates with $r=0$ corresponding to the center of the sample and $L$ is the system size. In the area where $0<\varphi_{i1}<\pi$ we have $\arg \Delta_{i2}<\arg \Delta_{i1}<\arg \Delta_{i3}$, while in the region $\pi<\varphi_{i1}<2\pi$ the phases of the order parameters have the opposite ordering $\arg \Delta_{i2}<\arg \Delta_{i3}<\arg \Delta_{i1}$. These two domains have the same phase ordering (up to global rotation) as the ones created by the initial conditions for the domain wall. For conventional single-flux-quanta vortex each gap was initialized as $\Delta_{i\alpha}=0.1\cdot\tanh\left(20r_i/L\right)\cdot e^{i\varphi_i + 2\pi\alpha/3}$. The vector potential is initially set to zero in all four cases.
Upon convergence, the domain wall is stable because of geometric pinning. To ensure that the vortex anzats-initiated configuration converges to a stable vortex solution, we choose a significantly large grid size relative to the solution size so that the interaction of the vortex with a boundary is smaller than the numerical grid pinning or numerical accuracy. 

\subsection{Vortex structure}\label{subsec:vort_str}

\begin{figure*}
\centering
\includegraphics[trim={1.5cm 0.5cm 1.5cm 0.5cm},clip,width=0.85\linewidth]{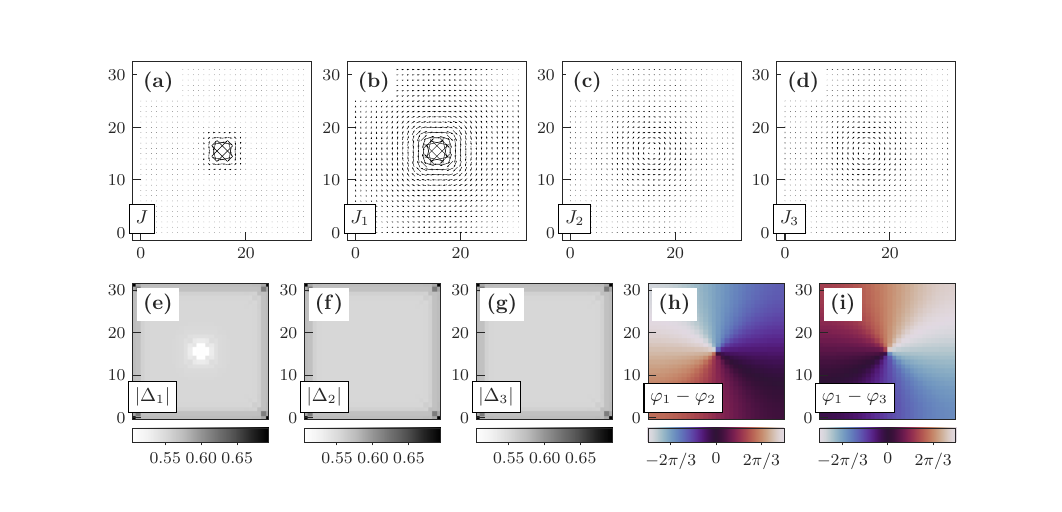}
\caption{Fractional vortex with phase winding only in the first component in a $U(1)^3$ three-component model. (a) Total current $J$ (\eqref{eq:curr}). (b-d) partial currents $J_1$, $J_2$, and $J_3$ (\eqref{eq:par_curr}). (e-g) absolute order parameter values $\Delta_1$, $\Delta_2$, and $\Delta_3$. (h) and (i) relative order parameter phases $\varphi_1-\varphi_2$ and $\varphi_1-\varphi_3$. The total magnetic flux in the system is $\Phi_0/3$. Square sample with linear size 32 with intraband coupling $V_{11}=V_{22}=V_{33} = 2.4$, zero interband coupling, $T=0.31T_c$ and $q=0.25$ is simulated.} \label{fig:fv_NJ}
\end{figure*}

\begin{figure*}
\centering
\includegraphics[trim={1.5cm 0.5cm 1.5cm 0.5cm},clip,width=0.85\linewidth]{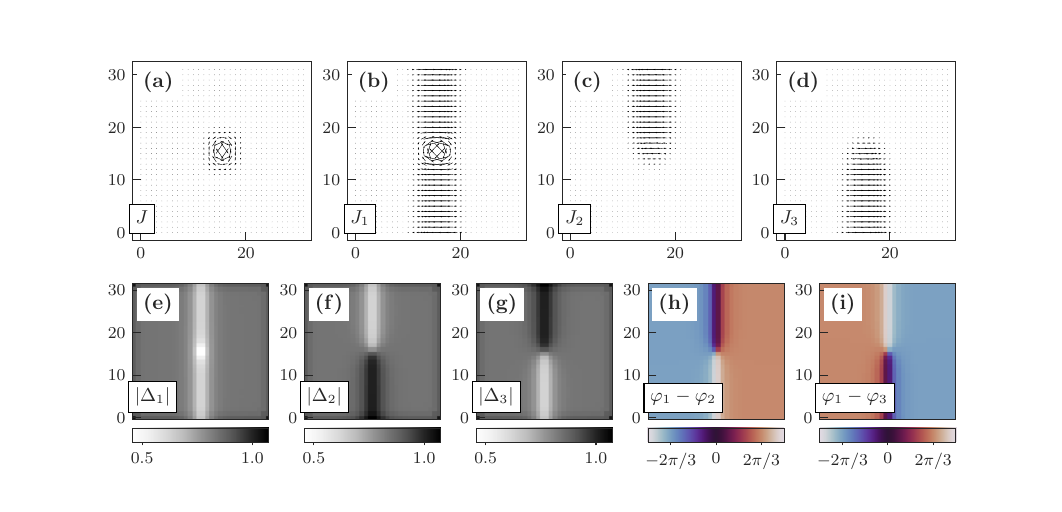}
\caption{Fractional vortex in a $U(1)\times Z_2$ symmetric three-component model. The total magnetic flux in the system is $\Phi_0/3$. Model parameters are the same, as in \figref{fig:fv_NJ}, except interband coupling $V_{12}=V_{13}=V_{23}=-0.6$ and temperature $T=0.27T_c$. On the domain wall, away from the vortex, partial currents compensate, and the total current is zero.} \label{fig:fv_JC}
\end{figure*}

There is a significant difference between the fractional vortex solutions for systems with (\figref{fig:fv_JC}) and without (\figref{fig:fv_NJ}) Josephson coupling. Let us define ``partial" current in band $\alpha$ as 

\begin{equation}\label{eq:par_curr}
    J_{ij\alpha} = -2q \sum_{\sigma} \Im{ \expval{c_{i\sigma\alpha}^\dagger c_{j \sigma \alpha}} \exp{\ii q A_{ij} }}.
\end{equation}

In a system without interband tunneling, we have
$U(1)^3$ case, with ``conserved" partial currents $J_\alpha$ (\figref{fig:fv_NJ}b-d). 
One can see from the solutions that the component with the phase winding (\figref{fig:fv_NJ}b) has a clockwise circulation of current. In this case, the intercomponent interaction is induced only by coupling to the vector potential.
The second and the third components (\figref{fig:fv_NJ}c,d) have counter-clockwise circulating currents because they do not have phase winding, and the current is due only to vector potential (note also the different magnitude of the current near the vortex core for components without phase winding).
That coexistence of clock- and anti-clock-wise circulation is closely connected with the fractionalization of flux quantum.
Another aspect that can be seen from the solutions is that the individual currents slowly decrease away from the vortex core, but the total current (\figref{fig:fv_NJ}a) is strongly localized. The oppositely circulating currents in different components far from the fractional vortex core lead to the divergence of the energy of the fractional vortex, as discussed in a different formalism based on the macroscopic London model in \cite{Babaev2002vortices}.

Nonzero Josephson coupling ($V_{\alpha, \beta} \ne 0$ if $\alpha \ne \beta$) causes direct intercomponent conversion, leading to several important effects. In the case of negative (repulsive) coupling, 
the phase differences have one of two stable values \figref{fig:fv_JC}h,i, except for the domain wall, where they have a non-optimal value in a narrow region. So, now, there are two different domains in the sample, separated by a domain wall. In this case, the energy of a vortex depends on the domain wall length and, hence, is proportional to the linear size of the sample. For a detailed discussion of the structure of domain walls at the level of phenomenological Ginzburg-Landau models, see \cite{garaud2011topological,garaud2013chiral,garaud2014domain}.

\subsection{Domain wall}\label{subsec:dw}

\begin{figure}
\centering
\includegraphics{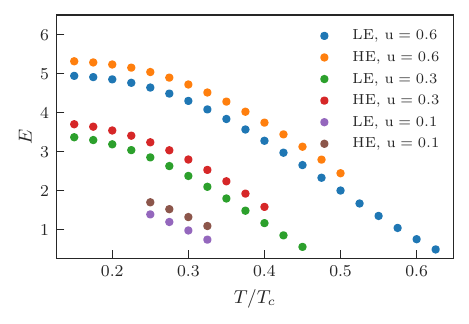}
\caption{Domain wall energy in bandwidth units as a function of temperature. A domain wall was generated in the middle of a square sample with $24\times24$ nodes. Results were obtained for slightly asymmetric intraband coupling $V_{11}=2.5$, $V_{22}=V_{33}=2.4$ and various interband couplings $V_{12}=V_{13}=V_{23}=-u$, $u=0.6,\,0.3,\,0.1$. The energy is calculated for both kinds of domain walls - high energy (HE) and low energy (LE). A characteristic temperature exists for each coupling strength, above which the HE domain wall becomes unstable on our numerical grid. For higher temperatures, the domain wall width becomes comparable to the system size.}
\label{fig:dw_energy}
\end{figure}

The \figref{fig:fv_JC} shows two different domain walls - the first above the vortex and the second below. Two gaps are suppressed on each of these two domain walls, and the third one is enhanced (\figref{fig:fv_JC}e-g).  If the bands are not symmetric, energetically different kinds of domain walls emerge. Such domain walls, in general, have different energies per unit length. As an example, we calculated domain wall energy $E=E_{\mathrm{DW}} - E_\mathrm{U}$ for several Josephson couplings as shown in \figref{fig:dw_energy}. Here $E_{\mathrm{DW}}$ stands for the total free energy \eqref{eq:Fsum} of a system with a domain wall, and $E_{\mathrm{U}}$ is the total free energy of an empty system. According to our results, if one of the gaps in the system has a larger amplitude, the domain wall, on which this gap is enhanced, has lower energy than a domain wall where this gap is suppressed. We call the first kind of domain wall a low energy (LE) domain wall and the second kind - a high energy (HE) domain wall.

A junction point between two different domain walls is a point-like topological defect, that is, as we see in Fig. \ref{fig:fv_JC}, a fractional vortex.
If these domain walls have different energies per unit length, 
the system minimizes its energy, reducing the length of the more energetically expensive domain wall. So, an effective force $F=\rho_\mathrm{HE} - \rho_\mathrm{LE}$ is acting on the vortex. Here $\rho_\mathrm{i} = E_i / L_i$ is the linear energy density, $E_i$ is the domain wall energy, and $L_i$ is the domain wall length.
This force makes vortices with fractional winding harder to stabilize by pinning, which may be related to the fact that fractional vortices so far were observed only for a particular value of doping.

However, it is possible to have a stable fractional vortex solution even if not all three bands are symmetric. If bands 2 and 3 have identical coupling constants and band 1 has different ones, both kinds of domain walls with suppressed band 1 still have identical energy density. Therefore, the fractional vortex between them is stable. The only qualitative difference between such an object and a fractional vortex in fully symmetric $U\left(1\right)\times Z_2$ is the non-unit fraction flux carried by this object (see Sec. \ref{subsec:B_field}).

Where several fractional vortices are connected with domain walls, they interact electromagnetically, by current-current interaction, by core-core interaction, and an effective force $F$, caused by the difference between the domain wall energy densities. 
When the repulsion is stronger than the attraction, the separate fractional vortices will be more energetically favorable than conventional ones, so the conventional vortex on top of the domain wall will split into fractional vortices despite being stable in a system without a domain wall. 
This is the case for the system in \figref{fig:fv_JC}, where we can observe an isolated fractional vortex on top of the domain wall. 

\begin{figure*}
\centering
\includegraphics[trim={1.5cm 0.5cm 1.5cm 0.5cm},clip,width=0.85\linewidth]{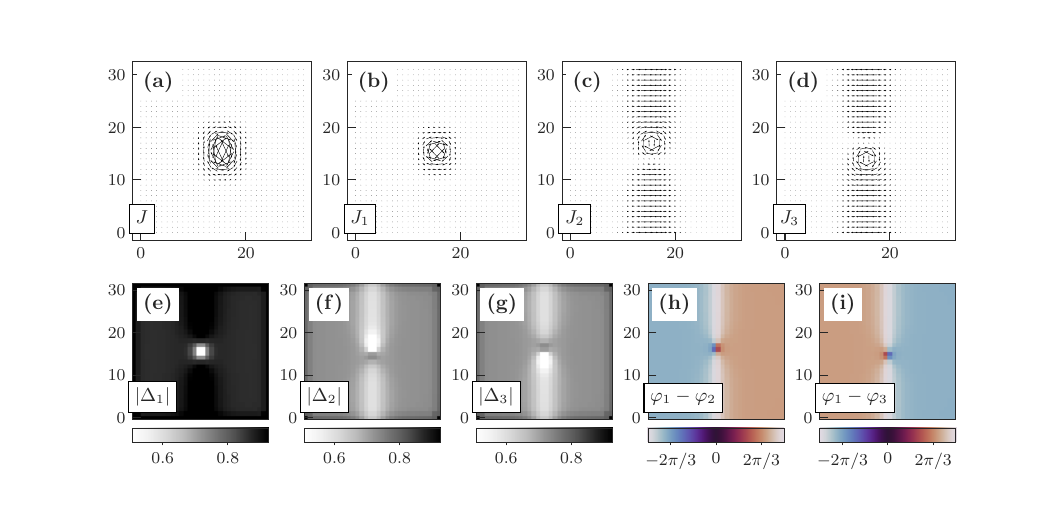}
\caption{Composite single-quantum vortex (consisting of very slightly offset fractional vortices) on top of the domain wall in a $U(1)\times Z_2$ asymmetric three-component model. The total magnetic flux in the system is $\Phi_0$. Model parameters are the same, as in \figref{fig:fv_NJ}, except interband coupling $V_{12}=V_{13}=-0.6$, $V_{23}=-0.5$ and temperature $T=039T_c$.}
\label{fig:fv_NS}
\end{figure*}

However, 
the opposite situation is also possible.
We demonstrate this parameter regime at \figref{fig:fv_NS}. We start the iteration procedure with only one fractional vortex in the first component. The domain walls attached to it are energetically unfavorable. The domain wall tension is strong enough to create two more fractional vortices on the boundaries of the sample and attract them to the original fractional vortex in the center of the sample. The resulting composite object has in total integer flux and consists of only slightly offset fractional vortices in all three components as can be seen from phase windings (\figref{fig:fv_NS}h,i). Therefore, it is possible to have rather compact composite objects on top of the domain walls, which in magnetometry will appear very similar to integer vortices.

\subsection{Tunneling conductance} \label{subsec:tunn_cond}
Scanning Tunneling Microscopy is a powerful technique for studying vortices, e.g. \cite{hess1990vortex,fischer2007scanning}. In our model, tunneling conductance may be calculated using \eqref{eq:tunn_cond}, allowing us to study the density-of-states signatures of domain walls and fractional vortices in $s+is$ superconductors.   

\begin{figure}
\centering
\includegraphics{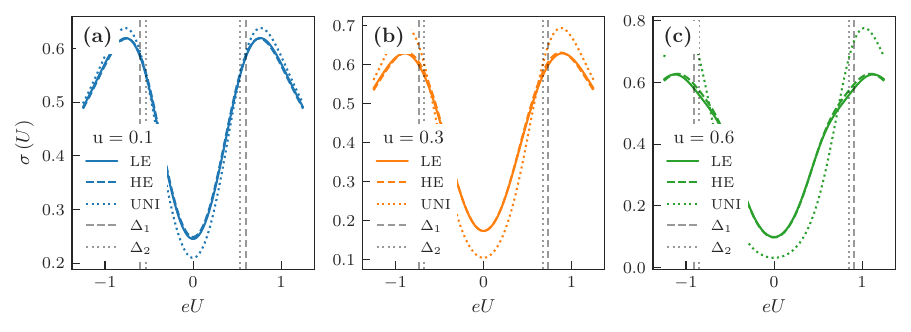}
\caption{Tunneling conductance \eqref{eq:tunn_cond} for different domain walls and uniform solution (UNI) in the absence of vortices. Simulation parameters are the same as on \figref{fig:dw_energy}, $T=0.25T_c$. For weak interband coupling $s+is$, the domain wall gives nearly no signature in tunneling conductance.}
\label{fig:dw_cond}
\end{figure}

\figref{fig:dw_cond} represents tunneling conductance curves for high energy (HE) and low energy (LE) domain walls, compared with tunneling conductance dependence in domain-wall-free samples. Domain walls for a certain coupling strength $u$ have noticeable suppression of the gaps and thus may be observed in STM. The domain wall conductance is higher than the bulk conductance below the superconducting gap and lower above the gap. It may also be seen that the patterns for both kinds of domain walls are pretty similar.

\begin{figure} 
\centering
\includegraphics{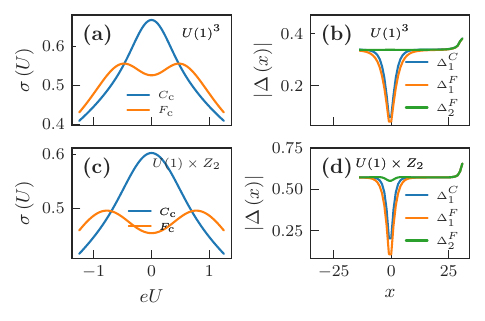}
\caption{(a) Tunneling conductance $\sigma\left(U\right)$ \eqref{eq:tunn_cond} for conventional $\left(C_\mathrm{c}\right)$ and fractional $\left(F_\mathrm{c}\right)$ vortex cores in $U(1)^3$ system. (b) Gap amplitude crossections for the system (a). (c) Tunneling conductance $\sigma\left(U\right)$ for conventional $\left(C_\mathrm{c}\right)$ and fractional $\left(F_\mathrm{c}\right)$ vortex cores in $U(1)\times Z_2$ system Fractional vortex is located between two different domain walls; there are no domain walls in the sample with the conventional vortex. (d) Gap amplitude crossections for the system (c).
In all four cases we analyze a square sample with a linear size of 64 nodes for a system with couplings $V_{11}=V_{22}=V_{33}=2.4$ and charge $q=0.25$. For $U(1)^3$ system we use $T=0.46T_c$ and zero intercomponent coupling; for $U(1)\times Z_2$ system we use $T=0.39T_c$ and repulsive intercomponent coupling $V_{12}=V_{13}=V_{23}=-0.6$.}
\label{fig:tun_cond}
\end{figure}

\figref{fig:tun_cond} demonstrates a significant difference between conventional and fractional vortices in $U(1)^3$ and $U(1)\times Z_2$ systems arising due to very different core structures. Fractional vortex has fewer electronic core states compared to the conventional vortex. Therefore, the tunneling conductance on the Fermi-level ($U=0$) for the fractional vortex is suppressed. So, Scanning Tunneling Microscopy can be used to distinguish conventional and fractional vortices in both systems with and without Josephson coupling \cite{newfract}.

\subsection{Magnetic field}\label{subsec:B_field}
 
In general, the shape and localization of the magnetic field and the total flux carried by the unconventional vortex in these models depend on the temperature and the strength of inter-component Josephson coupling. 

It was pointed out in the $U(1)\times U(1)$ Ginzburg-Landau models that general fractional vortices have power-law localization of the magnetic flux \cite{babaev2009magnetic}. Also, clear differences in the magnetic field distribution of fractional versus integer vortices were observed in the experiment  \cite{iguchi2023superconducting}. In the case of the Josephson-coupled system, the phase-difference mode is massive. Hence, we expect an exponential localization, but the overall localization is different.

\begin{figure} 
\centering
\includegraphics{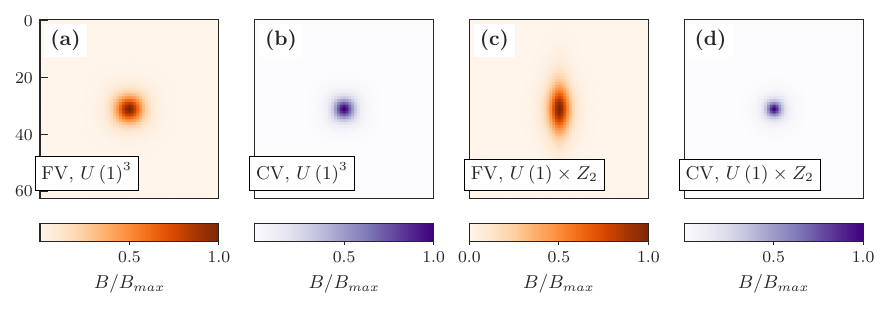}
\caption{(a), (b) - magnetic field for fractional (a) and conventional (b) vortices in $U(1)^3$ system. Both vortices have rotational symmetry. (c), (d) - magnetic field for fractional (c) and conventional (d) vortices in $U(1)\times Z_2$ system. 
 The  $U\left(1\right)\times Z_2$ fractional vortex is elongated along the domain wall. The model parameters for $U\left(1\right)^3$ and $U(1)\times Z_2$ cases are the same as on \figref{fig:tun_cond}.} \label{fig:B_field} 
\end{figure}

Concerning the shape of the magnetic field and the magnetic flux carried by the vortex, the following patterns can be observed - conventional vortex in both interacting and non-interacting systems is rotationally symmetric (apart from some corrections related to the square symmetry of our grid) and carries one quantum of magnetic flux \figref{fig:B_field}. The fractional vortex carries $1/3$ quanta in both cases due to the chosen symmetry between the components. However, in the presence of Josephson coupling, it elongates along the domain wall. 
According to \cite{Babaev2002vortices}, in the London model fractional vortex with winding in $i$-th component carries the flux $\Phi_i=\left|\Delta_i\right|^2/\sum_j\left|\Delta_j\right|^2 \Phi_0$. So, in a fully symmetric three-band model, we expect all three kinds of elementary vortices to carry $\Phi_0/3$ magnetic flux. However, in systems with non-symmetric components ($\left|\Delta_1\right|\ne\left|\Delta_2\right|$ or $\left|\Delta_1\right|\ne\left|\Delta_3\right|$), for example, in the systems on FIG. \ref{fig:dw_energy}, this is not the case, and a fractional vortex carries a non-unit fraction of flux quanta.

\subsection{Core states}\label{sec:core_states}
\begin{figure} 
\centering
\includegraphics{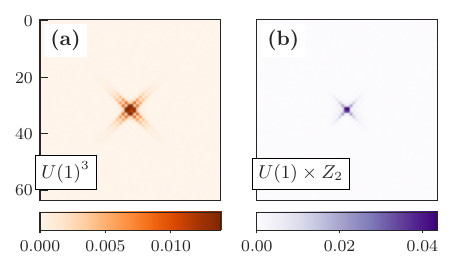}
\caption{Core state for fractional vortices with winding in only the first $\alpha=1$ band in $U(1)^3$ (a) and $U(1)\times Z_2$ (b) systems. Simulation parameters are the same as on \figref{fig:tun_cond}. Probability density $\left|u_{i 1 n}\right|^2$ is plotted for the electronic state with the lowest energy $E_{1i}$. The core states are well-localised in both systems.} \label{fig:core_states} 
\end{figure}

Similar to the ordinary vortices, due to the gap suppression in one of the bands, localized electronic states form in the cores of fractional vortices (FIG. \ref{fig:core_states}). Therefore, if a single electron occupies this state, the resulting composite object combines a fractional flux quanta tube and a well-localized charged particle. If several such objects are located far enough, their interaction may be neglected, and hence, these composite objects should obey anyonic statistics \cite{wilczek1982quantum}. This may be shown by the following simple example. We consider two identical fractional vortices carrying the flux $\Phi$ each. One of the core states in each of the vortices is occupied by an electron. If we move one of the vortices around the other one, the electron wavefunction will gain the Aharonov-Bohm phase  proportional to the ratio of the flux $\Phi$ to flux quantum $\Phi_0$ due to the interaction with the vector potential. Additional phase will be different from $2\pi$, since the vortices carry only a fraction of the flux quantum hence the objects demonstrate anyonic statistics.

\section{Discussion} \label{sec:concl}
The observation of vortices that carry a varying fraction of quantum flux (fractional vortices) in the multiband superconductor \bkfa requires a theoretical analysis of these objects. The nontrivial aspect of this material is the existence of only two broken symmetries, but at least three superconducting bands. According to the Ginzburg-Landau and London models-based arguments \cite{Babaev2002vortices,garaud2014domain}, this circumstance may lead to more than two fractional vortices. One cannot justify the existence of these objects by standard symmetry- and topology-based arguments; hence, microscopic justification is especially important.
This paper studied the solutions for fractional vortices and domain walls in a fully microscopic three-band Bogoliubov-de-Gennes model, including a fully self-consistent solution for magnetic fields.  The latter allows us to verify the existence of stable objects with magnetic flux quantum fractionalization. This also allows us to demonstrate the deviation of magnetic flux localization from that of ordinary vortices, manifested clearly manifested in the considered model in the fractional vortex elongation. The selected microscopic model provides us with direct access to the local density of states for conventional and fractional vortex solutions. This allows us to obtain specific tunneling conductance patterns for fractional vortex cores that might be used to distinguish fractional and conventional vortices. We find stable vortices whose magnetic field shape, localization, and density-of-states signatures significantly differ from those of the single-quanta vortices and can be directly probed in experiments. In the model of $s+is$ superconductors, we find that domain walls give a signature in the density of states. However, this signature is parameter-dependent and may be undetectable.
The experimental characterization of these vortices is also important because enclosing a variable fraction of the flux quantum makes them realize the charge-flux-tube bound state \cite{wilczek1982quantum}. Hence, these objects obey fractional statistics. 
Another application for these objects is fluxonics, which was previously based on conventional vortices \cite{Miyahara1985, Golod2015}.
 
\ack{
We thank A. Benfenati and M. Barkman for insightful discussions and for providing the code developed in \cite{iguchi2023superconducting}. 
}

\funding{This work was supported by the Swedish Research Council Grants  2022-04763, by Olle Engkvists Stiftelse, and the Wallenberg Initiative Materials Science for Sustainability (WISE) funded by the Knut and Alice Wallenberg Foundation.}

\roles{IT wrote the code, performed the investigation and calculations, analyzed  and interpreted the data, and wrote the paper.
EB supervised the project and contributed to the interpretation and writing the paper.}

\data{All data that support the findings of this study are included within the article.}

\bibliographystyle{iopart-num}
\bibliography{bibliography}

\end{document}